\documentclass[aps,nofootinbib,reprint,nobibnotes,showpacs,superscriptaddress]{revtex4-1}

\usepackage[english]{babel}

\usepackage{graphicx}

\usepackage{amsmath}
\usepackage{amssymb}
\usepackage[version=3]{mhchem}	
\usepackage{mathrsfs}
\usepackage{upgreek}

\usepackage[hidelinks,linktocpage]{hyperref}
	\hypersetup{colorlinks,linkcolor={red!70!black},citecolor={green!70!black},urlcolor={blue!70!black}}

\usepackage[usenames,dvipsnames]{xcolor}
\usepackage{soul}

\usepackage{comment}	

\newcommand{\imgFOLDER}{./}

\addto\captionsenglish{%
}

\newcommand{\bb}{$0\nu \beta \beta$} 
\newcommand{\bbvv}{$2\nu \beta \beta$} 

\newcommand{\TVLR}{$\text{TVL}^2+\text{TVR}^2$} 
\newcommand{\rejeff}{\varepsilon_\mathrm{rej}} 
\newcommand{\dt}{\Delta \text{t}} 

\begin{document}

	\title{Novel technique for the study of pile-up events in cryogenic bolometers}

	\author{A.~Armatol}
\affiliation{IRFU, CEA, Universit\'e Paris-Saclay, Saclay, France}

\author{E.~Armengaud}
\affiliation{IRFU, CEA, Universit\'e Paris-Saclay, Saclay, France}

\author{W.~Armstrong}
\affiliation{Argonne National Laboratory, Argonne, IL, USA}

\author{C.~Augier}
\affiliation{Institut de Physique des 2 Infinis, Lyon, France}

\author{F.~T.~Avignone~III}
\affiliation{University of South Carolina, Columbia, SC, USA}

\author{O.~Azzolini}
\affiliation{INFN Laboratori Nazionali di Legnaro, Legnaro, Italy}

\author{A.~Barabash}
\affiliation{National Research Centre Kurchatov Institute, Institute for Theoretical and Experimental Physics, Moscow, Russia}

\author{G.~Bari}
\affiliation{INFN Sezione di Bologna, Bologna, Italy}

\author{A.~Barresi}
\affiliation{INFN Sezione di Milano - Bicocca, Milano, Italy}
\affiliation{University of Milano - Bicocca, Milano, Italy}

\author{D.~Baudin}
\affiliation{IRFU, CEA, Universit\'e Paris-Saclay, Saclay, France}

% ...

\collaboration{CUPID Collaboration}
\noaffiliation

\date{\today}
	
	\begin{abstract}
		Precise characterization of detector time resolution is of crucial importance for next-generation cryogenic-bolometer experiments searching for neutrinoless double-beta decay,
		such as CUPID, in order to reject background due to pile-up of two-neutrino double-beta decay events.
		In this paper, we describe a technique developed to study the pile-up rejection capability of cryogenic bolometers.
		Our approach, which consists of producing controlled pile-up events with a programmable waveform generator, has the benefit that we can reliably and reproducibly control the
		time separation and relative energy of the individual components of the generated pile-up events. The resulting data allow us to optimize and benchmark analysis strategies to
		discriminate between individual and pile-up pulses.
		We describe a test of this technique performed with a small array of detectors at the Laboratori Nazionali del Gran Sasso, in Italy; we obtain a $90\%$ rejection efficiency 
		against pulser-generated pile-up events with rise time of $\sim 15$\,ms down to time separation between the individual events of about $2$\,ms.
		\\[9pt]
		Published on: \href{https://journals.aps.org/prc/abstract/10.1103/PhysRevC.104.015501}{Phys.\ Rev.\ C {\bf 104}, 015501 (2021)}

	\end{abstract}

	\maketitle

\section{Introduction}
\label{sec:intro}

	Two-neutrino double-beta decay (\bbvv), despite being a rare process, constitutes a dominant fraction of intrinsic radioactivity in low-background neutrinoless double-beta decay
	(\bb)~\cite{Furry:1939qr} detectors.
	Excellent detector energy resolution is essential to distinguish between {\bbvv} decays and {\bb} decay candidates. 
	However, for next-generation {\bb} searches, aiming at unprecedented low backgrounds, random pile-up of {\bbvv}~events can constitute a non-negligible continuum background in the region of 
	interest (ROI) around the \bb~transition Q-value~\cite{Chernyak:2012zz}.
	
	Cryogenic bolometers meet the requirements of excellent energy resolution, low background and large mass needed for a high-sensitivity {\bb} decay search.
	These detectors have set some of the most stringent limits on the {\bb} decay half-life in multiple isotopes, in particular 
	\ce{^{82}Se}~\cite{Azzolini:2019tta}, \ce{^{100}Mo}~\cite{CUPIDMo:2020}~and \ce{^{130}Te}~\cite{Alfonso:2015wka,Adams:2019jhp}.
	The CUORE Upgrade with Particle IDentification (CUPID), a proposed upgrade of the CUORE experiment, will search for {\bb} decay of \ce{^{100}Mo}, aiming at a half-life
	sensitivity greater than $10^{27}$\,yr~\cite{CUPIDInterestGroup:2019inu}.
	CUPID will consist of a large array of \ce{^{100}Mo}-enriched \ce{Li_2MoO_4} (LMO) scintillating bolometers instrumented with cryogenic light detectors (LDs).
	The experiment will exploit the experience acquired by running CUORE, the first tonne-scale cryogenic bolometer array, and will take advantage of its
	low-background cryogenic infrastructure~\cite{Alduino:2019xia};
	CUPID will combine this with the effective background reduction strategies
	demonstrated by the CUPID-0~\cite{Azzolini:2019nmi} and CUPID-Mo~\cite{Armengaud:2019loe} detectors to realize a one-tonne array with a background index of the order
	of $10^{-4}\,\text{counts\,keV}^{-1}\,\text{kg}^{-1}\,\text{yr}^{-1}$ in the ROI. 	

	In order to meet this goal, it is important to assess and mitigate each individual background component in the final technical design.
	Given the {\it short} decay half-life of \ce{^{100}Mo}, $7.1\times 10^{18}$\,yr~\cite{Armengaud:2019rll},
	the \bbvv~pile-up is expected to constitute a non-negligible fraction of the overall CUPID background budget.
	An estimate for a single CUPID-like crystal ($300$\,g of LMO with 100\% enrichment of $^{100}$Mo) assuming a 1-ms resolving time for the heat detector puts the expected
	background rate from pile-up of about $3.5\times 10^{-4}\,\text{counts\,keV}^{-1}\,\text{kg}^{-1}\,\text{yr}^{-1}$~\cite{CUPIDInterestGroup:2019inu}.
	Characterization of the pile-up induced background requires a thorough understanding of the resolving time CUPID-like bolometers can achieve. 
	
	CUPID will use neutron transmutation doped (NTD) thermistors~\cite{Larrabee:1984} as temperature (phonon) sensors.
	NTD-based macro-calorimeters have a relatively slow time response; for crystals of a few hundreds grams,
	the typical rise time of a thermal pulse is of the order of few tens of ms and the pulse decay is typically in the range of $(0.1-1)$\,s.
	In an ideal detector, the rise time is generally a function of the temperature-dependent NTD working resistance, while the decay time is proportional to the ratio of the
	absorber crystal heat capacity and the conductance to the thermal sink.
	Given the relatively slow response to an energy deposition, multiple particle interactions close in time will produce overlapping thermal pulses, resulting in a cumulative pulse,
	whose rising edge will be affected by the pile-up of the underlying events. Therefore, characterizing the rising edge of thermal pulses is crucial to discrimination of pile-up events.
	
	In this work, we describe a new technique we developed to study the time resolution of CUPID-like detectors, and produce a controlled sample of pile-up events to determine the
	pile-up rejection efficiency of a pulse shape analysis. We tested this technique in a dedicated measurement carried out at the Laboratori Nazionali del Gran Sasso (LNGS), in Italy. 

\section{Measurement}
\label{sec:measurement}

	An array of LMO crystals was deployed to study the performance of CUPID-like detectors, with crystals of similar dimension and shape to those that are intended to be used in the 
	actual experiment~\cite{HallC_paper}. This deployment is one of a series of bolometric tests planned to optimize the final technical design for CUPID.
	The array was operated at the Hall~C cryogenic facility at LNGS between the summer of 2019 and the spring of 2020.
	The final runs of this measurement were devoted to generating a controlled sample of pile-up events to study the pile-up rejection efficiency.
	
	The array consisted of 8 LMO cubic crystals ($45\times45\times45$\,mm$^3$) arranged in two 4-crystal floors.
	The crystals on the bottom floor were covered on the side with a reflecting aluminum foil, while those on the top floor were not. A total of 12 \ce{Ge} disks
	(diameter $44$-mm, thickness $0.175$\,mm)
	acting as LDs were placed below, in between, and above the crystal-floors, so that each LMO crystal faced two LDs. Each LMO crystal and LD was instrumented with a NTD sensor for the signal readout.
	For the pile-up study we focused on three LMO detectors which were instrumented with a functioning silicon heater, since pulsing the heater is crucial to generating the controlled
	sample of pile-up events.
	
	The detector was operated at about $18$\,mK and the measured NTD working resistances were in the range of $(10-50)$\,M$\Omega$. The observed rise time values for thermal pulses in
	the LMO crystals were approximately $15$\,ms.
	
	Detector signals were conditioned by custom front-end electronics boards, which include a programmable amplifier and anti-aliasing (Bessel-Thomson) filter~\cite{Arnaboldi:2017aek}.
	The filter cut-off frequency was set to $63$\,Hz, slightly higher than the expected signal bandwidth, to reduce the high frequency noise while not affecting the signal shape.
	The boards also provided independent programmable biases for the NTDs. The detector waveforms were acquired continuously with a sampling frequency of 2\,kHz. 
	The data stream was then triggered by means of an online derivative trigger~\cite{DiDomizio:2018ldc}. Each triggered pulse (i.\,e.\ an event) consists of a 5-s window opened
	around the trigger sample, which by construction is fixed to 1\,s from the start of the time window (pre-trigger).

	The runs dedicated to the study of pile-up background aimed at producing a controlled set of pile-up events on which to benchmark the rejection efficiency of our analysis algorithms.
	To do this, we continuously injected signal-like pulses in the detector, scanning a range of values for both the pulse amplitudes and the time separation between consecutive pulses.
	The pulses were induced by injecting a calibrated amount of energy (via Joule heating) into each crystal heater using a programmable waveform generator (Tektronix AFG1062).

	This technique is commonly used for the thermal gain stabilization of  bolometers~\cite{Alessandrello:1998bf,Carniti:2017zkr}.
	Our experience has shown that the heater can be used to correct the (temperature-dependent) detector gain against small temperature drifts because
	the pulse amplitude to temperature ratio behaves in the same way for heater pulses and particle pulses.
	However, in this work, for the first time we tried to emulate the shape of physics pulses with heater pulses, paying special attention to the rising edge.
	The underlying idea, now tested with a commercial module, is to include this feature in our pulser, thus being constantly able to monitor the pile-up discrimination of our detectors.
	The shape of physics pulses and heater-induced pulses can be different due to the mechanisms of phonon generation and propagation.
	A possible reason is that the heater should mainly inject acoustic phonons into the crystal, while particle interactions would produce optical phonons directly inside the absorber,
	which immediately begin to degrade~\cite{Levinson:1980}.
	The shape and amplitude of heater pulses might be affected by the heater-wire conductance, which appears as an additional component of the detector,
	and by the fact that in this case the energy is released always in a fixed spot on the surface of the absorber, unlike for physics events, which can produce energy deposits
	throughout the crystal volume. It is fair to point out that these hypotheses are still under investigation.
	
	Using the heater allows the detector to be thermally excited with precise control of the time separation and energy of the underlying excitation pulses.
	Given the goals of the study, this approach is more effective than using an intense radioactive source which has several disadvantages, including that the activity has to be tuned to a 
	suitable rate and the arrival time and energy of particles exciting the detector cannot be controlled on the event-by-event level. 

\begin{figure}[tb]
		\centering
		\includegraphics[width=.9\columnwidth]{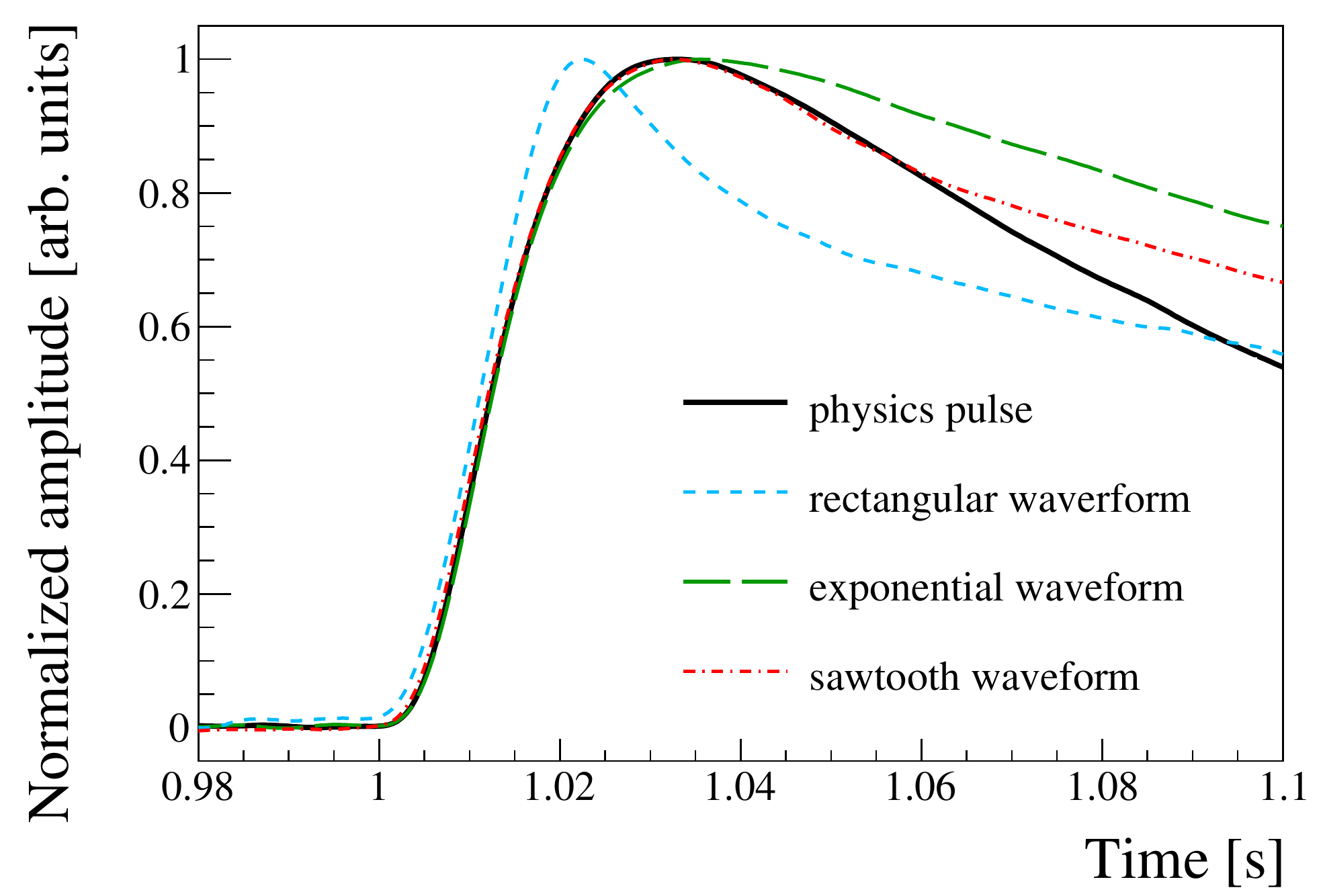}
		\caption{Overlay of the rising edge and initial part of the decay for heater pulses excited by rectangular (blue, en-dashed), exponential (green, em-dashed) and sawtooth (red, dot-dashed) waveforms
			and a particle pulse (black).
			Particle pulses have a similar shape, despite the different origin and entity of the energy deposition; quantitatively, the width of the distribution of
			rise and decay time is close to $3\%$.
		}
		\label{fig:pulse_shapes}
	\end{figure}

	The waveform generator was configured to deliver a particular class of pulses for a fixed amount of time.
	For each configuration, we set the number of pulses to be injected, the pulse amplitudes and the time interval between successive pulses.
	The shape of the input waveform was chosen so that the rising edge of the heater-induced pulses matched as closely as possible the rising edge observed for particle-induced pulses.
	Figure~\ref{fig:pulse_shapes} shows the rising edge observed for thermal pulses from particle events and heater events excited with rectangular, exponential and sawtooth waveforms.
	We found that the thermal pulses excited by sawtooth waveforms (right triangle with vertical rise and slow negative ramp) best reproduced the rising edge observed in particle induced pulses.

\begin{figure}[tb]
		\centering
		\includegraphics[width=.9\columnwidth]{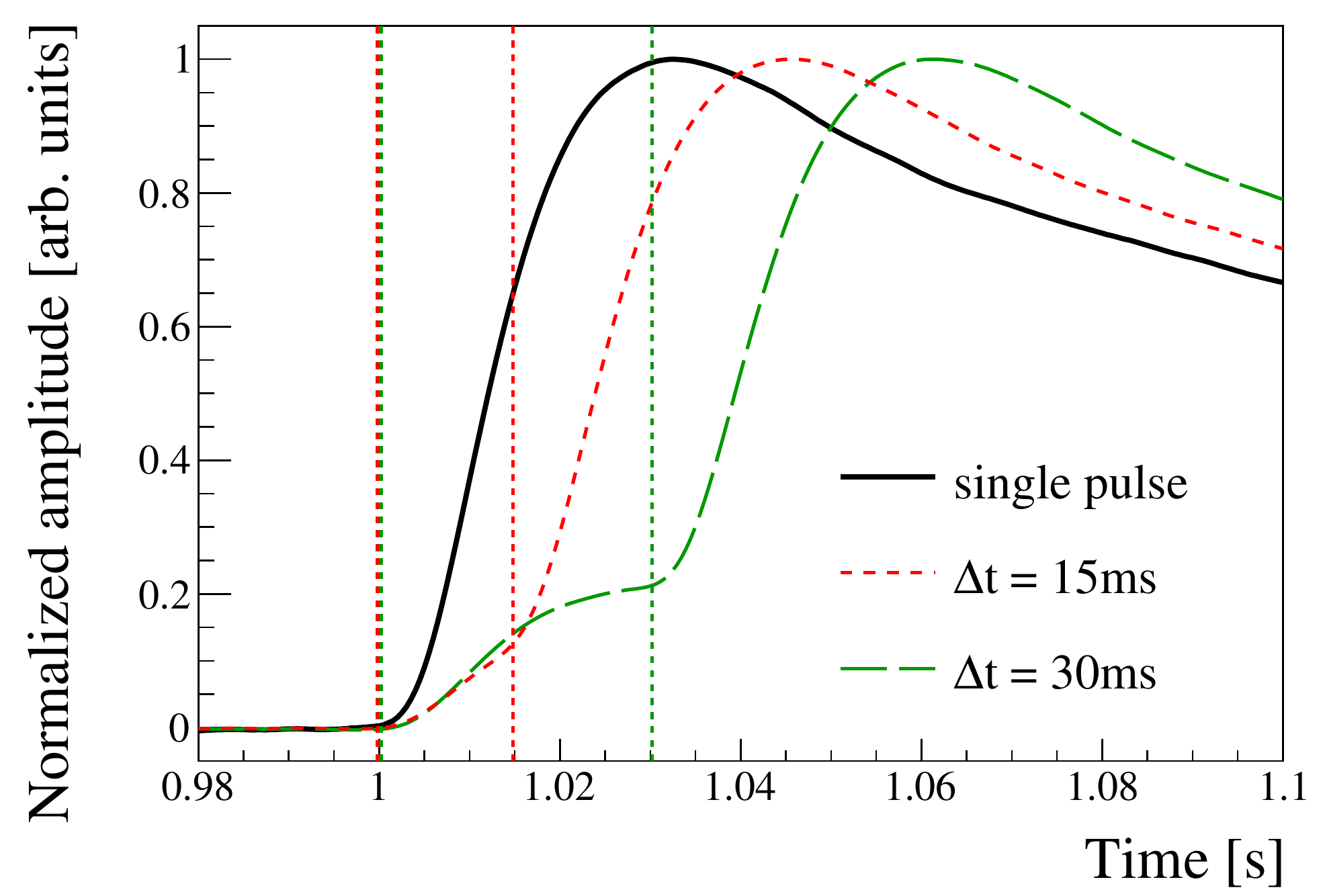}
		\caption{Examples of heater-induced pile-up pulses excited by two sawtooth waveforms with amplitude ratio $\alpha=1.4$ close in time.
			It can be seen that the time separations between the two pulses (the begin of the rise is taken as a reference) reflects that of the original waveforms ($\dt$).
			A time-isolated reference pulse is shown for comparison.}
		\label{fig:pileup_pulses}
	\end{figure}

	To generate reference individual pulses, the proxy for a well isolated particle events, we set the waveform-tooth width and amplitude to $50$\,ms and $170$\,mV, respectively,
	which produced pulses in the detectors with equivalent particle energies of $(1.5-2.5)$\,MeV on the different channels.
	The time interval between two successive individual waveforms was set to $15$\,s, so that the detectors had returned to baseline before the start of a new pulse.
	For the pile-up sample, we generated pairs of excitations with the second beginning before the detector had recovered from the first excitation in the pair.
	We set the waveform-tooth width to $50$\,ms for both excitations, fixed one amplitude to $170$\,mV and varied the other from $40$\,mV to $240$\,mV in steps of $50$\,mV.
	We denote the ratio of the two amplitudes in the pair by $\alpha$, and so this parameter varied between $0.24$ and $1.4$.
	We explored values of the time separation ($\dt$) between the pulse pairs in the pile-up sample ranging from $40$\,ms down to $1$\,ms (the Nyquist limit for the ADC sampling frequency was $1$\,kHz).
	Figure~\ref{fig:pileup_pulses} shows some of the observed thermal pulses; it can be seen that the distortion of the rising edge of pile-up pulses follows the time separation
	between the excitation pulses.
	
\section{Data analysis}
\label{sec:analysis}

	\begin{figure}[tb]
		\centering
		\includegraphics[width=1.\columnwidth]{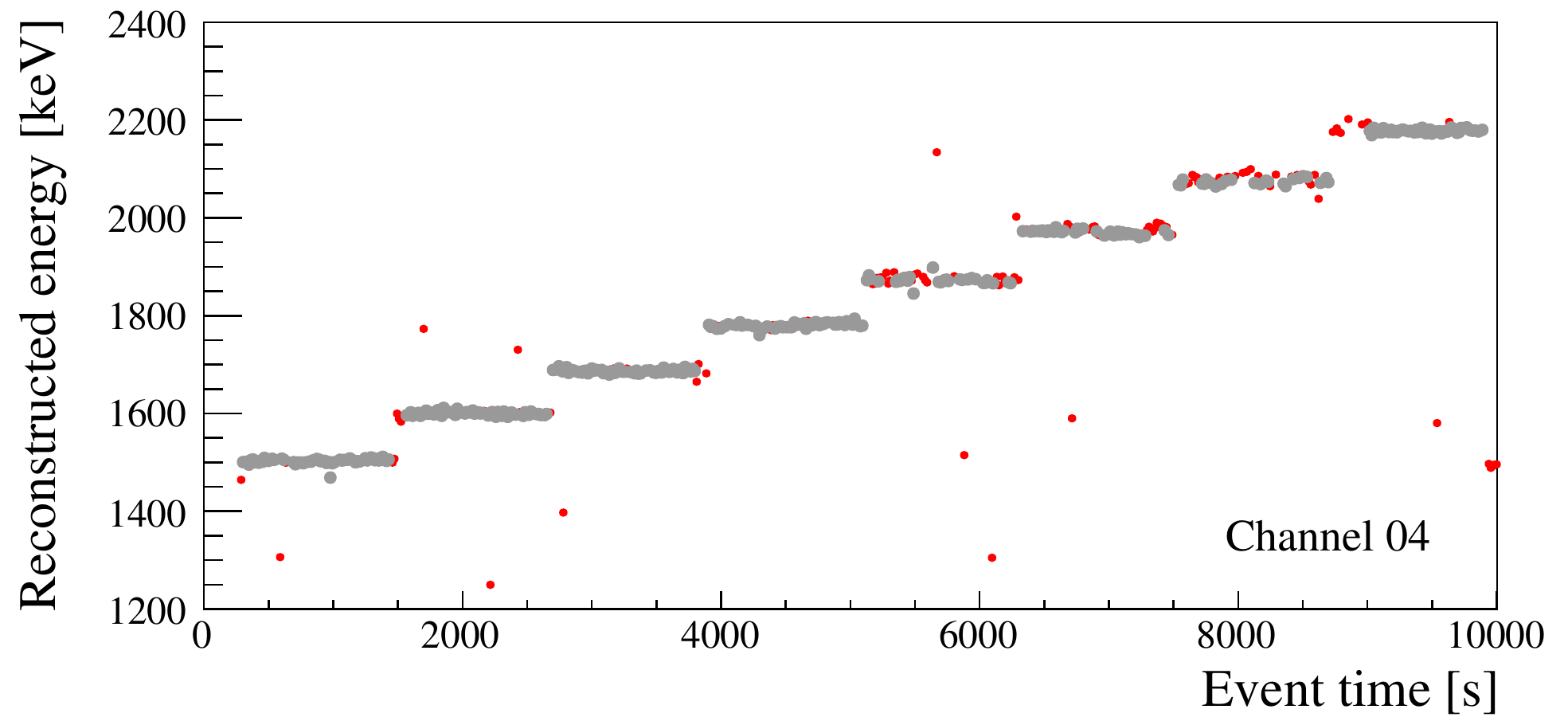}
		\caption{Time distribution of the events in one of the pile-up runs. Each interval corresponds to a different $\alpha/\dt$ configuration.
			In this sequence, $\alpha$ is fixed to $0.24$ while $\dt$ goes from $1$\,ms to $11$\,ms (from left to right).
			The gray dots represent the events selected as heater pulses, the red dots the discarded ones.
			To avoid an excessive statistics loss, we accepted events affected by a rather high noise level and by instability of the baseline, that would be otherwise discarded in a physics run.
			It is thus possible that some events in the various intervals deviate from the average behavior, despite the identical settings of the input waveforms.}
		\label{fig:intervals}
	\end{figure}

	The data analysis follows two main steps, a low-level processing and then a high-level analysis, i.\,e.\ the actual pile-up study.
	The low-level processing involves applying a software filter to optimize the signal-to-noise ratio and calculating a set of summary variables for each event such as pulse rise time,
	decay time and energy.
	These steps are analogous to the standard CUORE workflow~\cite{Alduino:2016zrl}, which has also been adopted as the starting point for the main analysis of the LMO detectors~\cite{HallC_paper}.
	These analysis tools are well established and have been applied over the years on multiple detectors~\cite{Andreotti:2010vj,Azzolini:2018yye}.
	The software filter uses the Optimum Filter (OF) technique~\cite{Gatti:1986cw}, which reduces the impact of noise on the reconstructed pulse amplitude and pulse shape parameters.
	In order to define the OF transfer function, we use the average of the time-isolated reference pulses as the signal response of the system, while we build the noise power spectrum
	from signal-less event windows acquired in the reference run.
	
	The first step of the high-level analysis is to associate each event with the corresponding waveform generator configuration. This is done based on the event \mbox{timestamp} since,
	as shown in Fig.~\ref{fig:intervals}, each waveform generator configuration corresponds to a specific time window. 
	
	We keep only clean events with a single trigger fired in the 5-s window. As pulser events dominate the event rate
	-- the random coincidence of pulser events and events due to natural radioactivity proved to be negligible during the measurement --
	we apply a $\pm 5$\,keV-cut around the median energy of the pulses for each configuration to suppress non-pulser events. 

	Once the events are selected for each configuration, we study how each of the pulse shape variables calculated in the low-level analysis are distributed for these events.
	We fit each distribution, one per channel, with a Gaussian function and take the resulting mean $M_{x,i}$ and standard deviation $\sigma_{x,i}$, where $x$ indexes the pulse shape variable studied
	(rise time, decay time, etc.) and $i$ indexes the waveform generator configuration or time interval.
	To quantify the utility of a variable to discriminate between individual isolated pulses and pile-up pulses, we define the {\it discrimination power} as the distance between the two
	corresponding distributions, namely
	\begin{equation}
		D \equiv \frac{\left| M_{x,i} - M_{x,R} \right|}{\sqrt{\sigma_{x,i}^2 + \sigma_{x,R}^2}},
		\label{eq:discr_power}
	\end{equation}
	where $R$ labels the (channel-dependent) reference quantities for individual isolated heater pulses.

	The pulse rise time and decay time are natural choices for the variables on which to discriminate since we expect the shape of the pulse to be different in case of isolated and pile-up pulses.
	However, we use a related combined variable based on the so-called {\it Test Value Left} (TVL) and {\it Test Value Right} (TVR).
	These are $\chi^2$-like parameters which quantify how well the shape of each filtered pulse matches the shape of the filtered average reference pulse on the left (i.\,e.\ rise) and
	right (i.\,e.\ decay) side of the pulse maximum.
	We find that the combination \TVLR~leads to an effective separation ($D>3$) already at values of $\dt$ of the order of a ms.
	Recalling Fig.~\ref{fig:pulse_shapes}, we note that this analysis is comparing heater-induced pile-up pulses to heater-induced isolated pulses and although we
	constructed the pulses to have similar rise times to particle induced pulses, the decay times are dissimilar.
	However, we expect the TVR variable will continue to provide some benefit when comparing isolated particle pulses to pile-up particle pulses as the decay time profile of a pile-up
	pulse is expected to be distorted compared to an isolated pulse. This will be tested in future studies.

\begin{figure}[tb]
		\centering
		\includegraphics[width=1.\columnwidth]{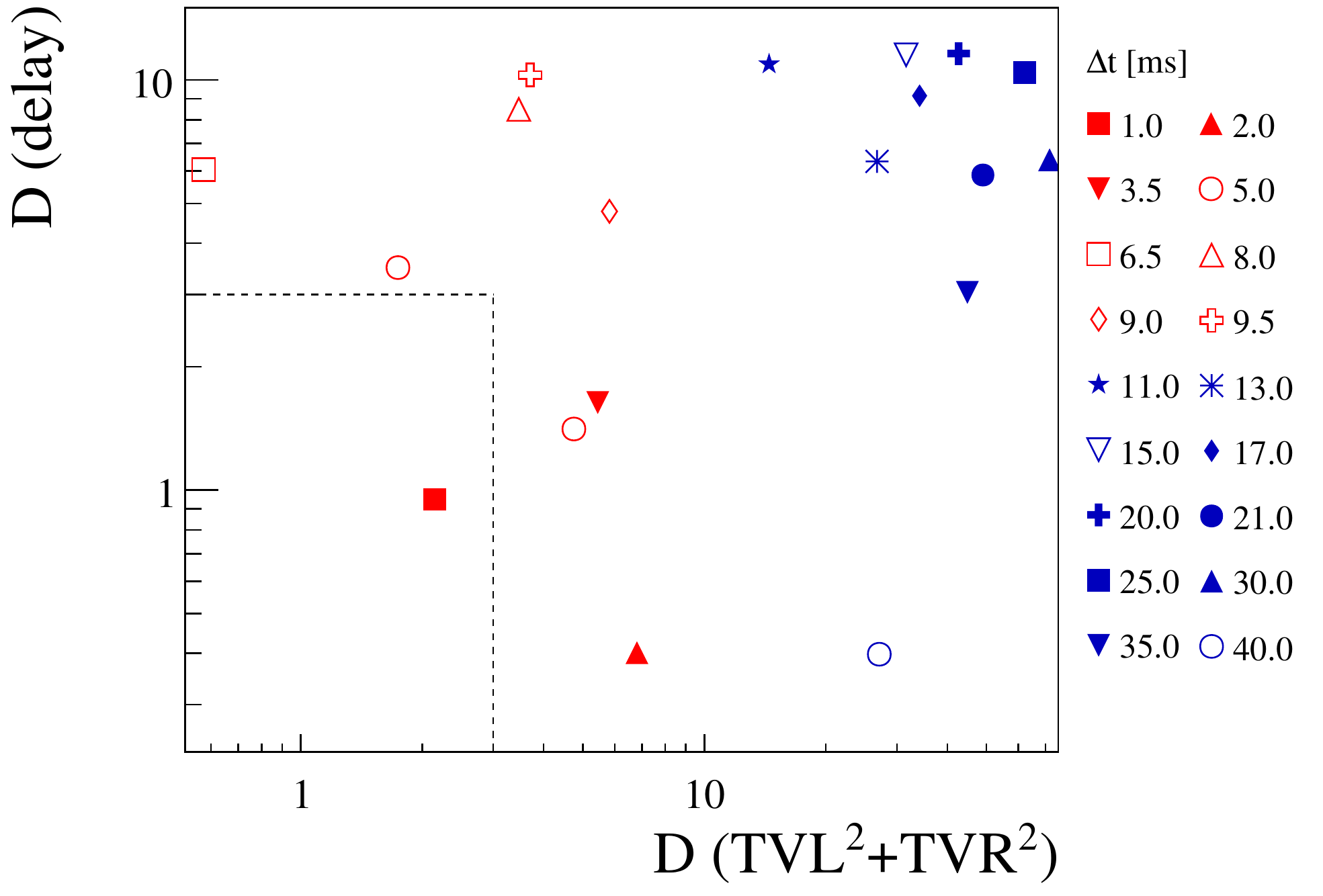}
		\caption{Comparison between the discrimination power ($D$) obtained with the \TVLR~and delay variables for the configuration $\alpha = 0.24$ of Channel\,03.
			The value $D=3$ for both variables is indicated with a dashed line.
			The configuration with $\dt=5$\,ms has been tested in two different runs; the different values of discrimination power are likely due to the different
			detector noise (see the discussion in the text).}
		\label{fig:discr_power}
	\end{figure}

	In addition to \TVLR, we consider another variable, the time separation between the start of the pulse window and the moment in which the filtered pulse reaches its maximum,
	which we call {\it delay}.
	The delay is sensitive to the noise and proved to be a good discriminator for some configurations of $\alpha$ and $\dt$ (Fig.~\ref{fig:discr_power}).
	Its complementarity to \TVLR~is due to the way we implement the OF.
	In order to calculate TVL and TVR, the maximum of the filtered pulse is aligned to that of the average pulse;
	the quality of the alignment depends on the residual noise of the filtered pulse.
	The more precisely the maxima are superimposed, the more the test values will be sensitive to the pulse distortions due to pile-up.
	On the other hand, if the alignment is not optimal, the \TVLR~might be misled by an overall matching of the pulses, but the delay variable would hint towards a different pulse
	shape due to possible pile-up.

\begin{figure}[tb]
		\centering
		\includegraphics[width=1.\columnwidth]{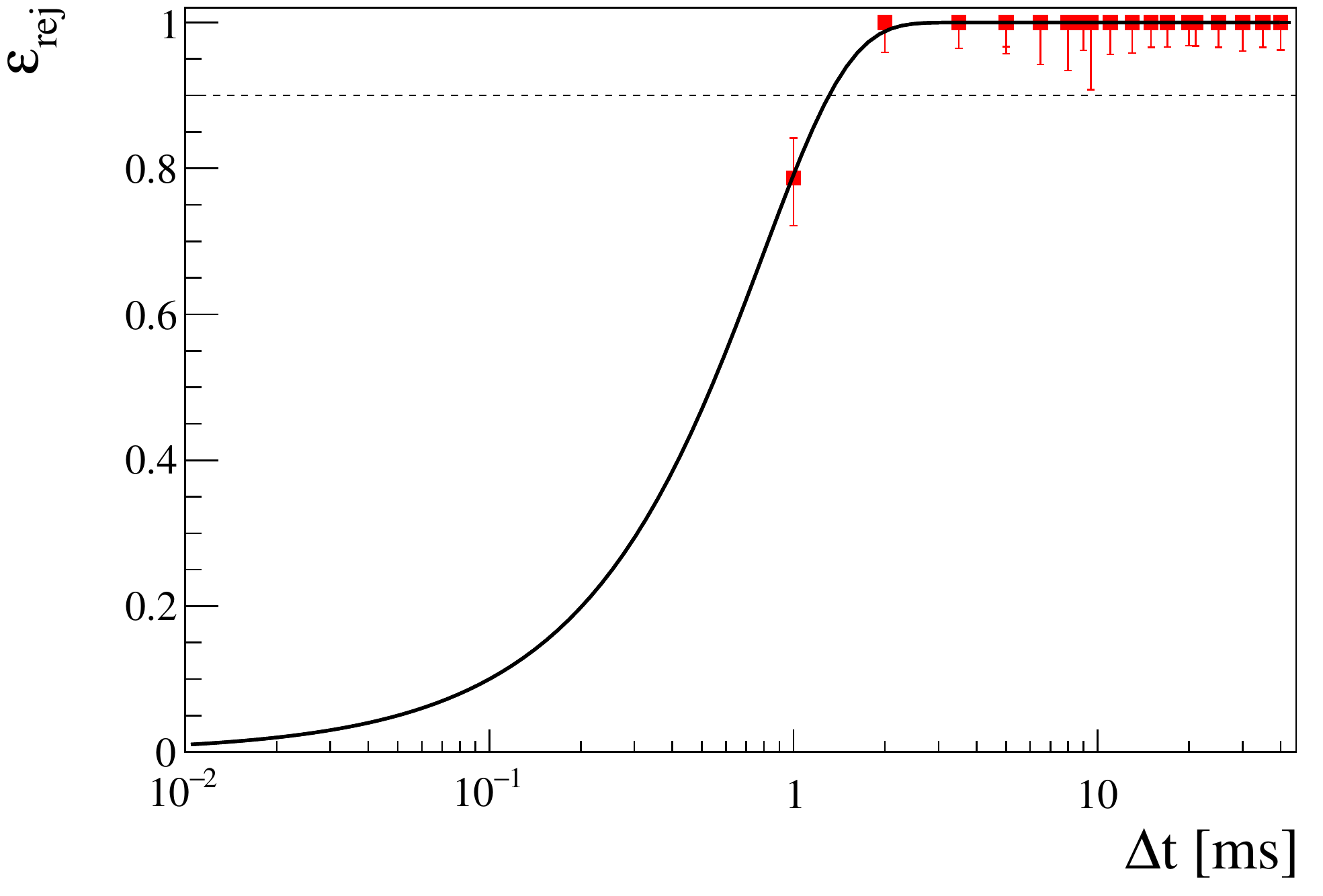}
		\caption{Pile-up rejection efficiency ($\rejeff$) as a function of $\dt$ for the configuration $\alpha = 0.24$ of Channel\,03.
			The solid line is the best-fit error function described in the text. The dashed horizontal line marks $\rejeff = 90\%$.}
		\label{fig:rej_eff}
	\end{figure}

	We analyze all the pulser events in the pile-up runs and define a {\it rejection efficiency}, $\rejeff$, as the fraction of events that lie outside the 
	$(M_{x,R} \pm 3\sigma_{x,R})$ range for at least one of the two considered variables, \TVLR~and delay.
	This interval is chosen so that the cut applied to the time-isolated reference pulses would select essentially all the events.
	For each $\alpha$ we fit the distribution of $\rejeff$ as a function of $\dt$ with an error function constrained to pass through the origin and extract the $\dt$ threshold
	corresponding to $\rejeff = 90\%$ (Fig.~\ref{fig:rej_eff}).

\begin{figure}[tb]
		\centering
		\includegraphics[width=1.\columnwidth]{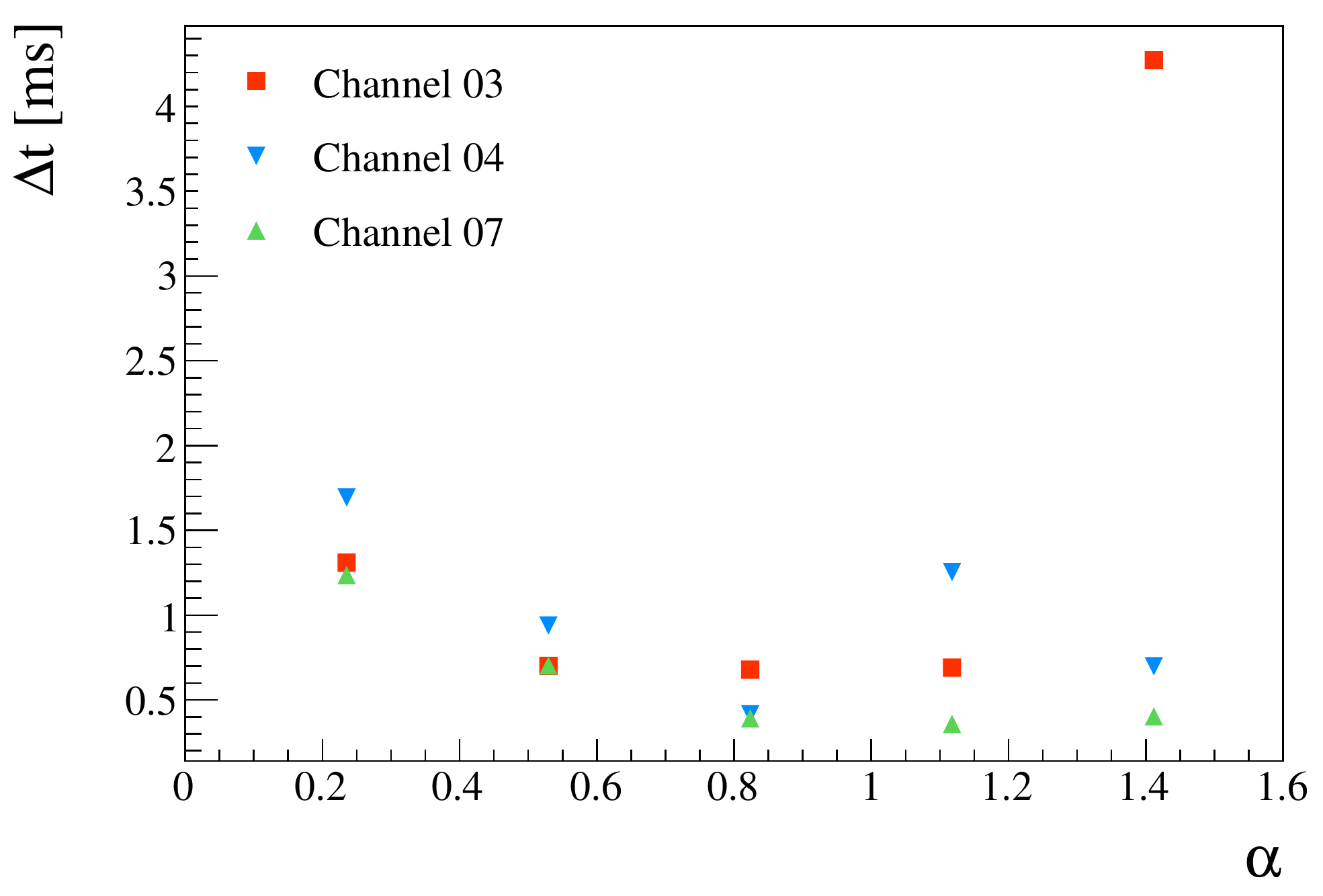}
		\caption{$\dt$ thresholds corresponding to $\rejeff = 90\%$ for each configuration of $\alpha$ obtained for the three channels.
			The two outliers at $\alpha=1.12$ for Channel\,04 and $\alpha=1.41$ for Channel\,03 are likely due to sub-optimal detector conditions while 
			collecting data for $\dt = 1$\,ms.} 
		\label{fig:dt_threshold}
	\end{figure}

	The combined results of this analysis are summarized in Fig.~\ref{fig:dt_threshold}, which shows the $\dt$ thresholds obtained for each $\alpha$ for each detector.
	The $90\%$ rejection efficiency is close to $2$\,ms or lower for most of the configurations. For completeness, we also performed the same analysis by considering the lone TVL instead
	of \TVLR, given the different decay time for particle and heater pulses, and noticed that the rejection efficiency worsens of about a factor two.
	The two outliers (corresponding to $\alpha=1.12$ for Channel\,04 and $\alpha=1.41$ for Channel\,03) are thought to be due to the sub-optimal detector operating conditions which
	we experienced while collecting the corresponding data.
	These points will be further investigated in future measurements; anyway they do not represent a major concern.
	We point out that the Hall-C cryogenic system allowed for daily measurements, while the cryostats hosting rare-event experiments, such as CUORE and CUPID-0,
	show month-long stability~\cite{Alduino:2016vjd,Alduino:2019xia}.
	Given the primary goal of testing our pulser-based technique, we preferred to acquire multiple and different configurations.
	This was achieved at the expense of statistics and of the optimal data quality (baseline stability and noise).
	During physics runs, sub-optimal operation conditions are identified and flagged, and the data rejected, to avoid a worsening of the overall performance.
	Data like those at the base of the two outliers of Fig.~\ref{fig:dt_threshold} would likely be discarded (still we deemed as correct to represent those points in the figure).

\section{Summary and outlook}
\label{sec:summary}

	We developed a new technique to investigate the ability of cryogenic bolometers to reject pile-up events. Using a controlled sample of pile-up signal events generated by
	injecting known excitations from a programmable waveform generator into the crystal heaters we identified a set of pulse shape parameters that allow for effective discrimination
	between individual and pile-up pulses, obtaining a $90\%$ rejection-efficiency for $\dt$ down to $2$\,ms for pulses with rise time of about $15$\,ms.
	A detailed Monte Carlo simulation of our detectors that incorporates this new pile-up rejection technique is still under development.
	However, our initial findings indicate that similar performance can be achieved for physics pulses.

	This work is part of a campaign of measurements planned to optimize the technical design of the CUPID detector and develop the analysis chain.
	In future measurements related to pile-up studies we plan to improve noise conditions and cryogenic stability, explore higher sampling frequencies to probe $\dt <1$\,ms and
	optimize the input excitation functions to better reproduce both the rising and falling edge of physics pulses.
	These data will be used to further benchmark the detector response simulations under development to assess the impact of pile-up on CUPID. 

	\begin{acknowledgments}
		The CUPID Collaboration thanks the directors and staff of the Laboratori Nazionali del Gran Sasso and the technical staff of our laboratories.
		This work was supported by the Istituto Nazionale di Fisica Nucleare (INFN); by the European Research Council (ERC) under the European Union Horizon 2020 program (H2020/2014-2020)
		with the ERC Advanced Grant No.\ 742345 (ERC-2016-ADG, project CROSS) and the Marie Sklodowska-Curie Grant Agreement No.\ 754496;
		by the Italian Ministry of University and Research (MIUR) through the grant Progetti di ricerca di Rilevante Interesse Nazionale (PRIN 2017, grant No.\ 2017FJZMCJ);
		by the US National Science Foundation under Grant Nos.\ NSF-PHY-1401832, NSF-PHY-1614611, and NSF-PHY-1913374.
		This material is also based upon work supported by the US Department of Energy (DOE) Office of Science under Contract Nos.\ DE-AC02-05CH11231 and DE-AC02-06CH11357; and
		by the DOE Office of Science, Office of Nuclear Physics under Contract Nos.\ DE-FG02-08ER41551, DE-SC0011091, DE-SC0012654, DE-SC0019316, DE-SC0019368, and DE-SC0020423.
		This work was also supported by the Russian Science Foundation under grant No.\ 18-12-00003 and the National Research Foundation of Ukraine under Grant No.\ 2020.02/0011.
		This research used resources of the National Energy Research Scientific Computing Center (NERSC).
		This work makes use of both the DIANA data analysis and APOLLO data acquisition software packages, which were developed by the CUORICINO, CUORE, LUCIFER and CUPID-0 Collaborations.
	\end{acknowledgments}

	
	\bibliography{ref}

\end{document}